\documentclass{article}
\usepackage{times,fancyhdr}
\usepackage{cite}
\usepackage{graphicx}
\usepackage{float}
\usepackage[utf8x]{inputenc}
\usepackage{subfig}
\usepackage{slashed}
\usepackage{multirow}

\usepackage{amssymb}
\usepackage{amsopn}
\usepackage{amsmath}
\usepackage{bm}
\usepackage{array}
\usepackage{cancel}
\usepackage{xcolor}
\usepackage{color}
\def\beq{\begin{equation}}
\def\eeq{\end{equation}}
\def\beqa{\begin{eqnarray}}
\def\eeqa{\end{eqnarray}}
\definecolor{dark}{rgb}{0.10,0.2,0.3}
\definecolor{light}{rgb}{1.7,1.5,0.6}
\definecolor{purpure}{rgb}{0.5,0.15,0.3}
\usepackage{hyperref}
\hypersetup{colorlinks,%
citecolor=dark,%
filecolor=dark,%
linkcolor=purpure,%
urlcolor=purpure,%
pdftex}

\title{Dijet Production at Large Rapidity Separation\\ in ${\cal N}=4$ SYM
\footnote{Preprint numbers: LPN11-30, IFT-UAM/CSIC-11-38, FTUAM-11-48}
} 
\author{M. Angioni$^1$, G. Chachamis$^2$, J.\,D. Madrigal$^3$, A. Sabio Vera$^{3,4}$ \\ 
\\
$^1$ Dipartimento di Fisica, Università di Firenze, Italy.\\
$^2$ Paul Scherrer Institut, CH-5232 Villigen PSI, Switzerland.\\
$^3$ Instituto de Física Teórica UAM/CSIC, Nicolás Cabrera 15,\\  \&
Universidad Autónoma de Madrid, E-28049 Madrid, Spain.\\
$^4$ Kavli Institute for Theoretical Physics,\\ University of California, Santa Barbara CA 93106, USA.}

\begin{document} 

\pagestyle{fancy}
\fancyhead{}
\fancyhead[EC]{M. Angioni, G. Chachamis, J.D. Madrigal, A. Sabio Vera}
\fancyhead[EL,OR]{\thepage}
\fancyhead[OC]{Dijet Production at Large Rapidity Separation in $\mathcal{N}=4$ SYM}
\fancyfoot{} 
\renewcommand\headrulewidth{0.5pt}
\addtolength{\headheight}{2pt} 

  \newcommand{\bra}[1]{\langle #1|}
\newcommand{\ket}[1]{|#1\rangle}
\newcommand{\braket}[2]{\langle #1|#2\rangle}
\newcommand{\opk}{\hat{\mathcal{K}}}
\newcommand{\asbar}{\bar\alpha_s}

\maketitle 

Ratios of azimuthal angle correlations between two jets produced at large rapidity separation are studied in the ${\cal N}=4$ super Yang-Mills theory (MSYM). It is shown that these observables, which directly prove the $SL(2,C)$ symmetry present in gauge theories in the Regge limit, exhibit an excellent perturbative convergence. They are compared to those calculated in QCD for different renormalization schemes concluding that the momentum-substraction (MOM) scheme with the Brodsky-Lepage-Mackenzie (BLM) scale-fixing procedure captures the bulk of the MSYM results.

\section{Introduction}

In recent years there have been many efforts to unveil the structure of the ${\cal N}=4$ super Yang-Mills (MSYM) theory, sometimes considered as ``the harmonic oscillator of the 21$^{\rm st}$ century"~\footnote{In a similar fashion, black holes can be thought of as the ``hydrogen atom" of quantum gravity~\cite{Maldacena:1996ky}.} since it seems to provide, in the planar limit, the first example of a solvable nontrivial quantum field theory in four dimensions. The discovery of an unexpected simplicity in the structure of scattering amplitudes~\cite{Parke:1986gb,Berends:1987me,Mangano:1990by,Witten:2003nn,Bern:2005iz} and the connection with string theory provided by the anti de Sitter / conformal field theory (AdS/CFT) duality~\cite{Maldacena:1997re,Gubser:1998bc,Witten:1998qj} support this reasoning.  

In the present work MSYM is taken as a theoretical laboratory to compare with observables calculated in 
Quantum Chromodynamics (QCD). This is useful since, for scattering amplitudes at high energies, both theories are equivalent at the leading order (LO) level, with the higher loop results in MSYM being, due to the enhanced symmetries and particle content,  much easier to compute. As a very interesting connection between both theories, it is known that the MSYM amplitudes, interpreted as part of QCD amplitudes and contributing to anomalous dimensions of gauge-invariant operators,  provide the ``highest degree of trascendentality" terms~\cite{Kotikov:2002ab}. This property is now a key ingredient to calculate anomalous dimensions to all-orders using 
integrability-based approaches~\cite{Beisert:2006ez}. 

Different observables have been proposed in the literature to be calculated in the context of MSYM~\cite{Hatta:2008st,Hatta:2007he,Levin:2009vj,Bork:2009nc}. The energy flow functions defined in terms of correlation functions of the energy-momentum tensor in~\cite{Hofman:2008ar} is a very interesting example. In QCD, typical observables are inclusive jet cross-sections. In the 
present work the study of ratios of azimuthal angle correlations in inclusive dijet production, for configurations where the two tagged jets are well separated in rapidity from each other, is carried out in detail. These ratios were first introduced in~\cite{Vera:2006un,Vera:2007kn}.

In \textsc{Section} \ref{2} the kinematical set-up under study is introduced and the calculation of the inclusive dijet production in the Balitsky-Fadin-Kuraev-Lipatov (BFKL) framework~\cite{BFKL1,BFKL2,BFKL3} at next-to-leading order~\cite{Fadin:1998py,Ciafaloni:1998gs} (NLO) both in QCD  (already known~\cite{Vera:2006un,Vera:2007kn,Colferai:2010wu}) and MSYM (a new calculation) in the minimal subtraction renormalization scheme ($\overline{\rm MS}$) is performed. In \textsc{Section} \ref{3} the momentum-subtraction (MOM) scheme~\cite{Celmaster:1979dm,Celmaster:1979km,Pascual:1980yu} with the Brodsky-Lepage-Mackenzie (BLM) scale-fixing procedure~\cite{Brodsky:1982gc} is applied to the  QCD calculation with the intention to challenge the common statement that it captures the bulk of the 
conformal contributions to all orders in the coupling. If this is correct it should then give a very similar 
result to that previously obtained in MSYM.  In \textsc{Section} \ref{4} it is shown that this is indeed the 
case, in particular, for the ratios of azimuthal angle correlations. Finally, the Conclusions and 
scope for future work are presented. 

\section{Azimuthal angle correlation ratios in MSYM 	\& QCD}\label{2}

In this section a review of the observables of interest is presented, following the notation in~\cite{Vera:2006un}. The 
kinematical  configuration under study is that of Mueller-Navelet jets\cite{Mueller:1986ey}, where two forward jets with very similar squared transverse momenta $\bm{p}_1^2$ and $\bm{p}_2^2$ are produced with a large rapidity separation $Y$ between them, and a relative azimuthal angle $\vartheta$. If $x_{1,2}$ are the fractions of longitudinal momentum from the parent hadrons carried by the partons generating the jets then $Y\simeq\ln{(x_1 x_2 s/\sqrt{\bm{p}_1^2\bm{p}_2^2})}$. In the limit of large rapidity separation, terms of the form $ \lambda Y$ must be resummed to all orders, with $\lambda$ being the 't Hooft coupling in MSYM 
and ${\bar \alpha}_s \equiv \alpha_s N_c / \pi$ in QCD. This is equivalent to the Regge limit, where 
$s \gg \sqrt{\bm{p}_1^2 \bm{p}_2^2}$, which can be treated using the  BFKL formalism.

In QCD, for the azimuthal angle correlation ratios to be defined at the end of this section, 
the effect of parton distribution functions is negligible at large $Y$, therefore the focus will be on partonic cross sections. 
These are written as the convolution of a differential partonic cross section with jet vertices 
$\Phi_{\rm jet_{1,2}}$:
\begin{equation}
\label{cross}
\hat{\sigma}(\bm{p}_1,\bm{p}_2,Y) =
\int d^2\bm{q}_1\int d^2\bm{q}_2 ~\Phi_{\rm jet_1} (\bm{q}_1, \bm{p}_1)\frac{d\hat{\sigma}}{d^2\bm{q}_1 d^2\bm{q}_2}\Phi_{\rm jet_2} (\bm{q}_2, \bm{p}_2),
\end{equation}
where, for simplicity, the LO jet vertex is used:
\begin{equation}
\Phi_{\rm jet_i}(\bm{q},  \bm{p}_i)\simeq \Phi^{(0)}_{\rm jet_i}(\bm{q},  \bm{p}_i)=\Theta (\bm{q}^2- \bm{p}_i^2).
\end{equation}
$ \bm{p}_i^2$ corresponds to the experimental resolution scale for the transverse momentum of the jet. Considering the Green function, $f$,  at  NLO, the gluon-gluon differential partonic cross section reads
\begin{equation}\label{diff}
\frac{d\hat{\sigma}}{d^2\bm{q}_1 d^2\bm{q}_2}=\frac{\pi^2 \bar{\alpha}_s^2}{2}\frac{f(\bm{q}_1, \bm{q}_2,Y)}{\bm{q}_1^2 \bm{q}_2^2}.
\end{equation}
Using the transformation
\begin{equation}
f(\bm{q}_1, \bm{q}_2,Y)=\int\frac{d\omega}{2\pi i}e^{\omega Y} \tilde{f}(\bm{q}_1, \bm{q}_2,\omega)
\end{equation}
the BFKL integral equation can be expressed in the form
\begin{equation}
\label{normalBFKL}
\omega \tilde{f} (\bm{q}_1, \bm{q}_2, \omega) = \delta^{(2)} (\bm{q}_1-\bm{q}_2) 
+\int d^2\bm{\kappa}~\mathcal{K}(\bm{q}_1, \bm{\kappa}) \tilde{f} (\bm{\kappa}, \bm{q}_2,\omega).
\end{equation}
At LO accuracy the solution to this equation is
\begin{equation}\label{LL}
\tilde{f} (\bm{q}_1, \bm{q}_2,\omega)=\sum_{n=-\infty}^\infty \int_{-\infty}^\infty 
\frac{d\nu}{2\pi^2} \frac{(\bm{q}_1^2/\bm{q}_2^2)^{i\nu}}{\sqrt{\bm{q}_1^2\bm{q}_2^2}}\frac{e^{in(\vartheta_1-\vartheta_2)}}{\omega-\bar{\alpha}_s\chi_0(|n|, \nu)},
\end{equation}
where the eigenvalue of the kernel is a function of the logarithmic derivative of the Euler Gamma function:
\begin{eqnarray}
\chi_0(n, \nu) =2\psi (1)-\psi\left(\frac{1+n}{2}+i\nu\right)-\psi\left(\frac{1+n}{2}-i\nu\right).
\end{eqnarray}
It is convenient to use the following transverse momenta operator notation:
\begin{equation}
\hat{q}|\bm{q}_i\rangle=\bm{q}_i|\bm{q}_i\rangle;\quad \langle\bm{q}_1|\hat{\mathbf{1}}|\bm{q}_2\rangle=\delta^{(2)}(\bm{q}_1-\bm{q}_2),
\end{equation}
to introduce the basis of eigenfunctions $|n, \nu\rangle$ of the LO kernel that appear in eq.~\eqref{LL}, satisfying the normalization 
condition $\langle n', \nu'| n, \nu\rangle=\delta (\nu-\nu')\delta_{n,n'}$, as

\begin{equation}\label{basis}
\langle \bm{q}|n, \nu\rangle =\frac{1}{\pi\sqrt{2}}(\bm{q}^2)^{i\nu-\frac{1}{2}}e^{in\vartheta}.
\end{equation}
At NLO, in the QCD case, this basis is no longer diagonal due to the scale invariance breaking stemming from  the running of the coupling contributions. The action of the NLO QCD kernel on the basis of eq.~\eqref{basis} reads~\cite{Kotikov:2002ab}
\begin{equation}
\begin{aligned}
\label{kernelnlo}
  &\bra{n,\nu}\opk\ket{\nu',n'} = \bar{\alpha}_{s, \overline{\rm MS}}
  \bigg[\chi_0\left(|n'|,\frac{1}{2}+i\nu'\right)\left(1-\frac{\bar{\alpha}_{s, \overline{\rm MS}}\beta_0}{8N_c}
  \left(i\frac{\partial}{\partial\nu}-i\frac{\partial}{\partial\nu'}-2\ln\mu^2\right) \right)\\
&\hspace{0.2cm}+\bar{\alpha}_{s, \overline{\rm MS}} \chi_1\left(|n'|,\frac{1}{2}+i\nu'\right) 
+ i \frac{\bar{\alpha}_{s, \overline{\rm MS}}\beta_0}{8N_c} \left(\frac{\partial}{\partial \nu'}\chi_0\left(|n'|,\frac{1}{2}+i\nu'\right)\right)
\bigg]\delta_{n,n'}\delta(\nu-\nu'),
\end{aligned}
\end{equation}
where a symmetrized version in $\nu$ and $\nu'$ to express $\ln \bm{q}^2$ as a derivative with respect to either $\nu$ or $\nu'$ has been used. This is the natural choice in the $\bm{q}_1^2 \simeq \bm{q}_2^2$ case, removing terms of the form $\chi_0'$ in the kernel, which would be complex for real values of $\nu$. All the other functions in eq.~\eqref{kernelnlo}, with $\nu = i \left(\frac{1}{2}-\gamma\right)$, are
\begin{equation}\label{eq1}
\begin{aligned}
\chi_1(n,\gamma) & =\mathcal{S}\chi_0(n,\gamma)+\frac{3}{2}\zeta(3)-\frac{\beta_0}{8N_c}\chi_0^2(n,\gamma)
+{\rm \Omega}(n,\gamma)-\frac{\pi^2\cos(\pi\gamma)}{4\sin^2(\pi\gamma)(1-2\gamma)}\\
&\hspace{-1cm}\times\left[\left( 3+\left(1+\frac{N_f}{N_c^3}\right)\frac{2+3\gamma(1-\gamma)}{(3-2\gamma)(1+2\gamma)}\right)\delta_{n,0}-\left(1+\frac{N_f}{N_c^3}\right)\frac{\gamma(1-\gamma)}{2(3-2\gamma)(1+2\gamma)}\delta_{n,2}\right];
\end{aligned}
\end{equation}
\begin{equation}
\begin{aligned}\label{omega}
{\rm \Omega}(n,\gamma) & = \frac{1}{4}\left[\psi''\left(\gamma+\frac{n}{2}\right)+\psi''\left(1-\gamma+\frac{n}{2}\right)\right. -2\phi(n,\gamma)-2\phi(n,1-\gamma)\Big];
\end{aligned}
\end{equation}
\begin{equation}
\begin{aligned}
\phi(n,\gamma)&=\sum_{k=0}^{\infty}\frac{(-1)^{k+1}}{k+\gamma+\frac{n}{2}}\Bigg(\psi'(k+n+1)-\psi'(k+1)\\&+(-1)^{k+1}(\beta'(k+n+1)+\beta'(k+1))+\frac{\psi(k+1)-\psi(k+n+1)}{k+\gamma+\frac{n}{2}}\Bigg);
\end{aligned}
\end{equation} 
\begin{eqnarray}
4 \,\beta'(\gamma) = \psi'\left(\frac{1+\gamma}{2}\right)-\psi'\left(\frac{\gamma}{2}\right), \quad
\mathcal{S}  =\frac{1}{12}\left(4-\pi^2+\frac{5\beta_0}{N_c}\right), \quad \beta_0=\frac{11}{3}N_c-\frac{2}{3}N_f.
\end{eqnarray}

In MSYM the absence of running of the coupling makes the action of the NLO kernel on the LO eigenfunctions diagonal and 
eq.~\eqref{kernelnlo} simplifies to
\begin{eqnarray}
\bra{n,\nu}\opk_{\rm MSYM}\ket{\nu',n'} 
= \lambda \left( \chi_0\left(|n'|,\frac{1}{2}+i\nu'\right)+\lambda \chi_1^{\rm MSYM}\left(|n'|,\frac{1}{2}+i\nu'\right) \right) 
\delta_{n,n'}\delta(\nu-\nu'),
\end{eqnarray}
with $\lambda$ being the non-running analogue of $\asbar$, and~\cite{Kotikov:2002ab}~\footnote{The  
MSYM calculations in this study have been obtained within the $\overline{\rm MS}$ scheme, other renormalization approaches give very similar results.}
\begin{equation}
\chi_1^{\rm MSYM}(n, \gamma)=\frac{\left(1-\zeta(2)\right)}{12}\chi_0(|n|,\gamma)+\frac{3}{2}\zeta (3)+{\rm \Omega} (|n|, \gamma).
\end{equation}

Expressed in terms of the $|n, \nu\rangle$ basis, the differential cross section in the azimuthal angle 
$\phi=\vartheta_1-\vartheta_2-\pi$, with $\vartheta_i$ corresponding to each tagged jet, is~\cite{Vera:2006un}
\begin{equation}
\frac{d\hat{\sigma}({\bf p}_{1,2}^2,Y)}{d\phi}=\frac{\pi^2\bar{\alpha}_s^2}{4\sqrt{{\bf p}_1^2 {\bf p}_2^2}}\sum_{n=-\infty}^\infty e^{in\phi}\mathcal{C}_n(Y),
\end{equation}
with the coefficients $\mathcal{C}_n (Y)$ given, in the QCD case, by
\begin{equation}\label{Cn}
{\cal C}_n \left(Y\right) =
\int_{-\infty}^\infty \frac{d \nu}{2 \pi}\frac{e^{{\bar \alpha}_s \left({\bf p}^2\right) Y \left(\chi_0\left(\left|n\right|,\nu\right)+{\bar \alpha}_s  \left({\bf p}^2\right) \left(\chi_1\left(\left|n\right|,\nu\right)-\frac{\beta_0}{8 N_c} \frac{\chi_0\left(\left|n\right|,\nu\right)}{\left(\frac{1}{4}+\nu^2\right)}\right)\right)}}{\frac{1}{4}+\nu^2},
\end{equation}
where, for simplicity, the configuration ${\bf p}_1^2 \simeq {\bf p}_2^2 \simeq {\bf p}^2$ has been taken. 
In the MSYM case, with $\beta_0 =0 $, these are
\begin{equation}\label{CnMSYM}
{\cal C}_n \left(Y\right) =
\int_{-\infty}^\infty \frac{d \nu}{2 \pi}\frac{e^{\lambda Y \left(\chi_0\left(\left|n\right|,\nu\right)
+ \lambda \chi_1^{\rm MSYM} \left(\left|n\right|,\nu\right)\right)}}{\frac{1}{4}+\nu^2}.
\end{equation}
Interesting observables can be obtained from these coefficients. ${\cal C}_0$ generates the total cross section 
since it is integrated over $\phi$:
\begin{equation}\label{total}
\hat{\sigma}({\bf p}_{1,2}^2,Y)=\frac{\pi^3 \bar{\alpha}_s^2}{2\sqrt{{\bf p}_1^2 {\bf p}_2^2}}\mathcal{C}_0(Y).
\end{equation}
To study the effect of the higher conformal spins $n$, which is one of the main points of this work~\footnote{The identification of the $n$'s as conformal spins stems from the elastic amplitude, linked to the total cross section via the optical theorem, which has an underlying $SL(2,C)$ invariance\cite{Lipatov:1985uk}.}, the following ratios of azimuthal angle correlations are introduced:
\begin{eqnarray} 
\label{cosmphi}
\langle \cos (m\phi)\rangle &=& \frac{\mathcal{C}_m (Y)}{\mathcal{C}_0 (Y)}, \\
\label{ratios}
{\cal R}_{m,n} (Y) &\equiv& \frac{\langle \cos (m\phi)\rangle}{\langle \cos (n\phi)\rangle} ~=~\frac{\mathcal{C}_m(Y)}{\mathcal{C}_n(Y)}.
\end{eqnarray}
In the analysis of these quantities different types of coefficients $\mathcal{C}_n$ will be used. The MSYM coefficients given in eq.~\eqref{CnMSYM} will be denoted with $\mathcal{C}_n^{\rm MSYM}$. For QCD, when computed with the full expression as in eq.~\eqref{Cn}, $\mathcal{C}_n^{\rm NLL}$ will be used, keeping $\mathcal{C}_n^{\rm LL}$ to indicate the use of the LO approximation in the exponential. Another useful quantity will be the scale invariant coefficient $\mathcal{C}_n^{\rm SI}$ given by putting $\beta_0=0$ in eq.~\eqref{Cn}. The 't Hooft limit, defined by $N_c\to\infty$ while keeping the product $\alpha_s N_c$ fixed, will be also investigated.

\section{Physical renormalization schemes \& BLM procedure}\label{3}

After the computation of the NLO terms in the BFKL kernel~\cite{Fadin:1998py,Ciafaloni:1998gs}, it was noticed that these corrections lead to instabilities~\cite{Ross:1998xw} in collinear regions. All-order resummations of the leading collinear contributions helped improve the perturbative convergence~\cite{Salam:1998tj,Vera:2005jt}. Other proposals 
based on introducing short-range in rapidity correlations were also successful in stabilizing the perturbative 
expansion\cite{Schmidt:1999mz,Forshaw:1999xm}. A complementary approach is that developed 
in~\cite{Brodsky:1998kn,Brodsky:2002ka} which, instead of employing the $\overline{\rm MS}$ scheme with arbitrary renormalization scale setting, proposed the use of a physical renormalization scheme like MOM~\cite{Celmaster:1979dm,Celmaster:1979km} with optimal scale set by the BLM procedure~\cite{Brodsky:1982gc}. 
In this way the NLO corrections have a milder behaviour and the value of the pomeron intercept has a very weak dependence on the hard scale of the scattering process, leading to an approximate scale invariance. This is a good motivation to compare this approach with MSYM, which enjoys four dimensional conformal invariance. 

The transition from the $\overline{\mbox{MS}}$ to the MOM scheme is equivalent, at LO accuracy, to the 
redefinition of the coupling~\cite{Celmaster:1979dm,Celmaster:1979km}
\begin{equation}
\alpha_{\text{MOM}}=\alpha_{\overline{\text{MS}}}
\left(1+T_{\text{MOM}}\frac{\alpha_{\overline{\text{MS}}}}{\pi}\right),
\end{equation}
where $T_{\text{MOM}}$ is a function of $N_c$, $N_f$ and of the gauge parameter $\xi$:
\begin{equation}
\begin{aligned}
T_{\text{\text{MOM}}}& =  T_{\text{\text{MOM}}}^{\text{conf}}+T_{\text{\text{MOM}}}^{\beta}, \\
T_{\text{\text{MOM}}}^{\text{conf}} &= \frac{N_c}{8} \biggl( \frac{17}{2} I +
\xi \frac{3}{2} (I-1) + \xi^2 \left(1-\frac{1}{3}I \right) -\xi^3 \frac{1}{6} \biggr)  ,\\
T_{\text{\text{MOM}}}^{\beta} &=- \frac{ \beta_0}{2} \biggl(1 +\frac{2}{3} I \bigg) ,
\end{aligned}
\end{equation}
with $I=-2 \int^{1}_{0}dx \ln(x)/[x^2-x+1]\simeq 2.3439$. At NLO this redefinition is equivalent to the rescaling 
\begin{equation}\label{res1}
\mu\to\bar{\mu}=\mu\exp \left(-\frac{T_{\text{MOM}}}{2\beta_0}\right).
\end{equation}
An appropriate choice of renormalization scheme and scale can render the higher order coefficients of the perturbative series small. Several prescriptions to do so have been proposed~\cite{Grunberg:1980ja,Stevenson:1980du,Stevenson:1981vj,Grunberg:1982fw}, with possibly the best physically motivated being the BLM one 
where the coupling redefinition absorbs charge renormalization corrections  in such a way that the coefficients of the perturbative series are identical to those of the corresponding conformally invariant theory with $\beta=0$.

In~\cite{Brodsky:1998kn} the BLM approach was applied to the BFKL description of the $\gamma^* \gamma^*$ 
cross section. In order to enhance the effect of BLM in gluon dominated processes, it was argued that it is appropriate to use a physical scheme suitable for non-abelian interactions, 
such as MOM, based on the 3-gluon vertex~\cite{Celmaster:1979dm,Celmaster:1979km,Pascual:1980yu}  or the $\Upsilon$ scheme based on $\Upsilon\to ggg$ decay. This procedure is not free from complications since it has an unnaturally high scale at $\nu=0$~\cite{Thorne:1999rb}.

In the present work the BLM setting for dijet production is explored, not only at the pomeron intercept level but, very importantly, also investigating its effects in azimuthal angle decorrelations. To generalize the BLM scale fixing for arbitrary conformal spins $n$ it is convenient to write the BFKL kernel in the form
\begin{equation}\label{kernelact}
\omega _{\overline{\text{MS}}}(\bm{q}^{2}, n, \nu ) =  \chi_0 (n, \nu ) \frac{\alpha_{\overline{\text{MS}}}(\bm{q}^{2}) N_c \,}{\pi } \Biggl( 1 + r_{\overline{\text{MS}}}(n, \nu )
\frac{\alpha_{\overline{\text{MS}}}(\bm{q}^{2})}{\pi } \Biggr).
\end{equation}
The  NLO coefficient $r_{\overline{\text{MS}}}$ is decomposed into $\beta$-dependent and conformal ($\beta$-independent) parts:
\begin{equation}
r_{ \overline{\text{MS}}} (n, \nu)\ =\ r_{\overline{\text{MS}}}^{\beta}(n, \nu)
+ r_{ \overline{\text{MS}}}^{\text{conf}} (n, \nu)\ ,
\label{evnl}
\end{equation}
where
\begin{equation}
r_{\overline{\rm MS}}^{\beta}(n, \nu)=-\frac{\beta_0}{4}\left(\frac{\chi_0(n,\nu)}{2}-\frac{5}{3}\right),
\end{equation}
and
\begin{eqnarray}
r_{\overline{\rm MS}}^{\text{conf}}(n, \nu) &=& -\frac{N_c}{4\chi_0(n,\nu)}\Bigg\{\frac{(\pi^2-4)}{3}\chi_0(n,\nu)-6\zeta(3)+\frac{\pi^2}{2\nu}\text{sech}(\pi\nu)\tanh (\pi\nu)\nonumber\\
&& \hspace{-2cm}-\Bigg[\psi''\left(\frac{n+1}{2}+i\nu\right)+\psi''\left(\frac{n+1}{2}-i\nu\right)
-2 \phi\left(n,\frac{1}{2}+i\nu\right)- 2 \phi\left(n,\frac{1}{2}-i\nu\right)\Bigg]\\
&&\hspace{-2cm}\times\Bigg[\Bigg(3+\left(1+\frac{N_f}{N_c^3}\right)\left(\frac{3}{4}-\frac{1}{16(1+\nu^2)}\right)\Bigg)\delta_{n,0}-\left(1+\frac{N_f}{N_c^3}\right)\left(\frac{1}{8}-\frac{3}{32(1+\nu^2)}\right)\delta_{n,2}\Bigg]\Bigg\}. \nonumber
\end{eqnarray}
The NLO BFKL intercept in the MOM-scheme, evaluated at the optimal BLM scale can be represented as follows:
\begin{equation}\label{inter}
\omega^{\text{MOM}}(\bm{q}^{2\, \text{MOM}}_{~~\text{BLM}},n,\nu) =
\frac{\alpha_{\overline{\text{MS}}}(\bm{q}^{2\, \text{MOM}}_{~~\text{BLM}})N_c}{\pi} \chi_{0} (n,\nu) \Biggl(1 +
r^{\text{MOM}} (n,\nu) \frac{\alpha_{\overline{\text{MS}}}(\bm{q}^{2\, \text{MOM}}_{~~\text{BLM}})}{\pi} \Biggr) ,
\end{equation}
where $r_{\text{MOM}}(n, \nu)  =  r_{\overline{\text{MS}}} (n, \nu) + T_{\text{MOM}}$. The BLM redefinition of the coupling involves the $\beta$-dependent part of the corrections implying that
\begin{equation}\label{qblmn}
\bm{q}^{2\, \text{MOM}}_{~~\text{BLM}} (n, \nu) = \bm{q}^2 \exp
\Biggl( - \frac{4 r^{\beta}_{\rm MOM}(n,\nu)}{\beta_0} \Biggr)
= \bm{q}^2 \exp \Biggl( \frac{1}{2}\chi_0 (n, \nu) + \frac{1+4 I}{3} \Biggr), 
\end{equation}
with $r^\beta_{\text{MOM}}(n, \nu)  =  r^\beta_{\overline{\text{MS}}} (n, \nu) + T_{\text{MOM}}$.

\section{Comparing MSYM with QCD results}\label{4} 

\begin{figure}[htbp]
\centering
\includegraphics[scale=0.8]{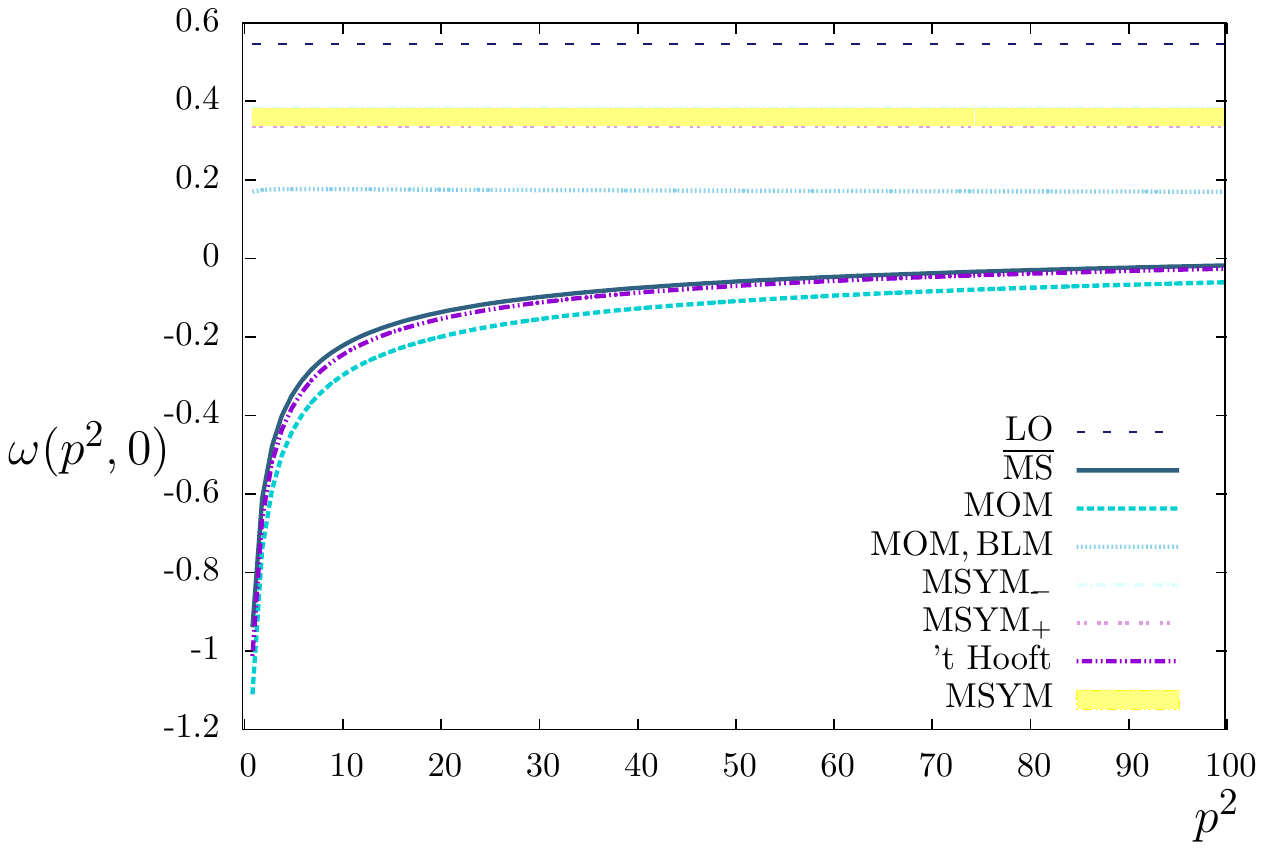}
\caption{Comparison of the evolution of the intercept with the value of the jet resolution scale ${\bf p}^2$ in LO and NLO for different renormalization schemes in QCD and the MSYM case. The curve $\omega_{\rm 't\,Hooft} ({\bf p}^2)$ corresponds to taking the 't Hooft limit in the QCD NLO case.}
\label{beta}
\end{figure}

As a first result, in \textsc{Fig.} \ref{beta} the intercept of eq.~\eqref{inter} for $n=0$ computed in the MOM scheme with BLM scale is confronted with the MSYM intercept and the results at LO and NLO with no BLM scale fixing. The coupling for MSYM is chosen in a range between $\lambda=\bar{\alpha}_s ({\bf q}^2/4)$ (MSYM$_-$) and $\bar{\alpha}_s (4 {\bf q}^2)$ (MSYM$_+$), which corresponds to the light yellow band. This plot is in agreement with the calculation performed for the $\gamma^* \gamma^*$ total cross section in~\cite{Brodsky:1998kn}
since the scale invariance, with respect to the photon virtualities in that case and the jet transverse momentum now, of the asymptotic intercept in the MOM scheme is manifest. This intercept for the conformal invariant MSYM theory at NLO is very close to the LO one, already hinting towards a better perturbative convergence than in QCD. For completeness, results for the QCD calculation taking the 't Hooft limit $N_c \to \infty$ are also shown. It is 
important to note that the MOM-BLM is the closest to MSYM of all renormalization schemes in QCD, in explicit agreement with the idea of it being collecting the conformal contributions to the observable. 

\begin{figure}[htbp]
\centering
\includegraphics[scale=0.8]{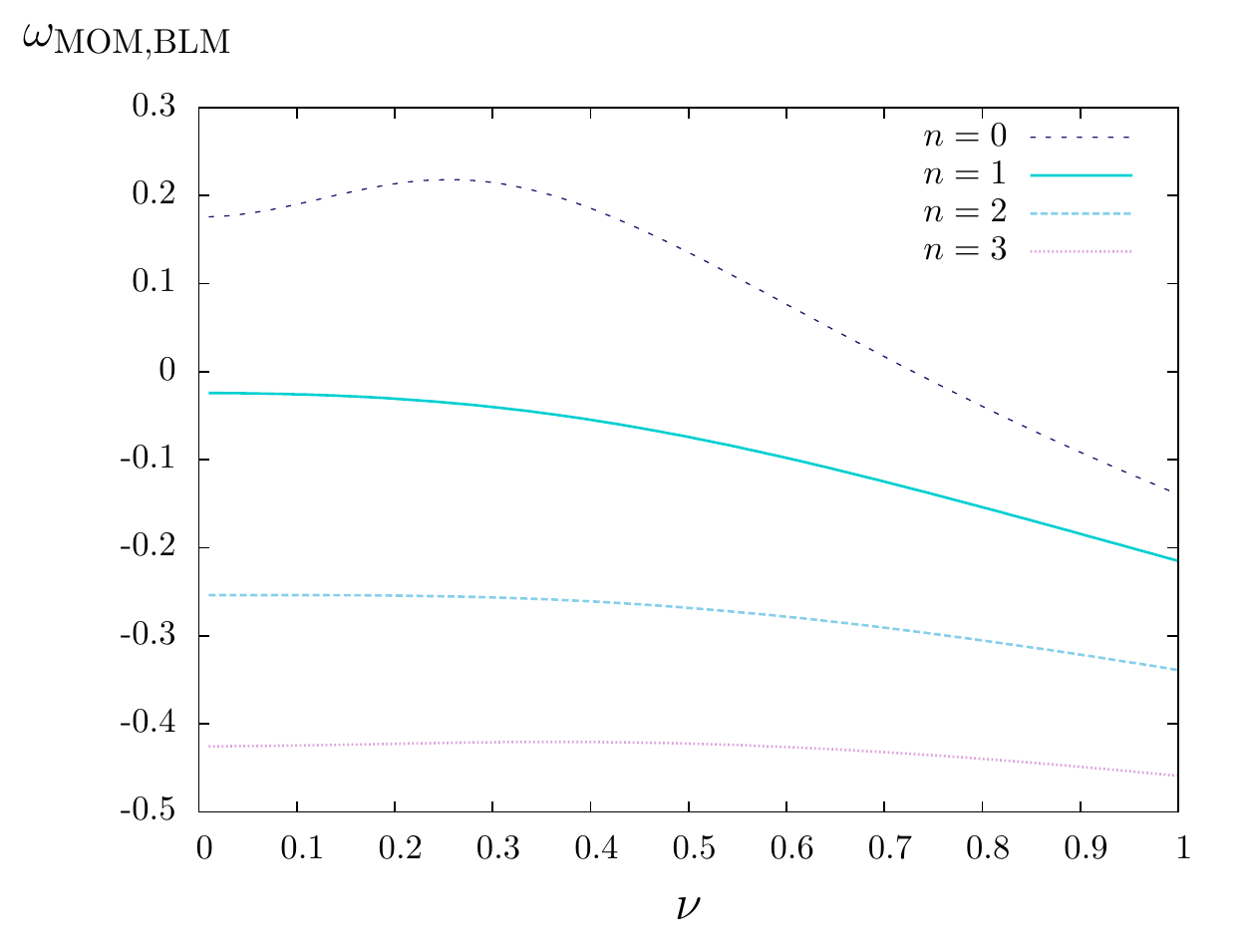}
\caption{Eigenvalues of the QCD BFKL kernel at NLO in MOM-BLM scheme, for different conformal spins $n$.}
\label{gamma}
\end{figure}

In \textsc{Fig.} \ref{gamma} a typical scale ${\bf p}^2=15~{\rm GeV}^2$ is taken to show, for QCD, the asymptotic 
intercepts for each of the azimuthal angle Fourier components $n$ in the MOM-BLM scheme, which can be read from the values of the curves at $\nu=0$. The dominant component at large relative rapidity separation of the dijets is $n=0$ with all the other intercepts being negative. A similar behaviour is found for other schemes and in the MSYM case. In the following it will be investigated how these intercepts affect physical 
observables sensitive to the azimuthal angle dependence.

As it was discussed in previous sections, the azimuthal angle dependence of the gluon Green function can be studied by expanding it in Fourier 
components, each of them characterized by a discrete parameter $n$, which in elastic scattering has the 
interpretation of a conformal spin associated to an underlying $SL(2,C)$ symmetry. In this way, by studying observables sensitive to the azimuthal angle, the Mo{\" e}bius invariance present 
in gauge theories at high energies is under scrutiny. 
\begin{figure}[htbp]
\centering
\includegraphics[scale=0.8]{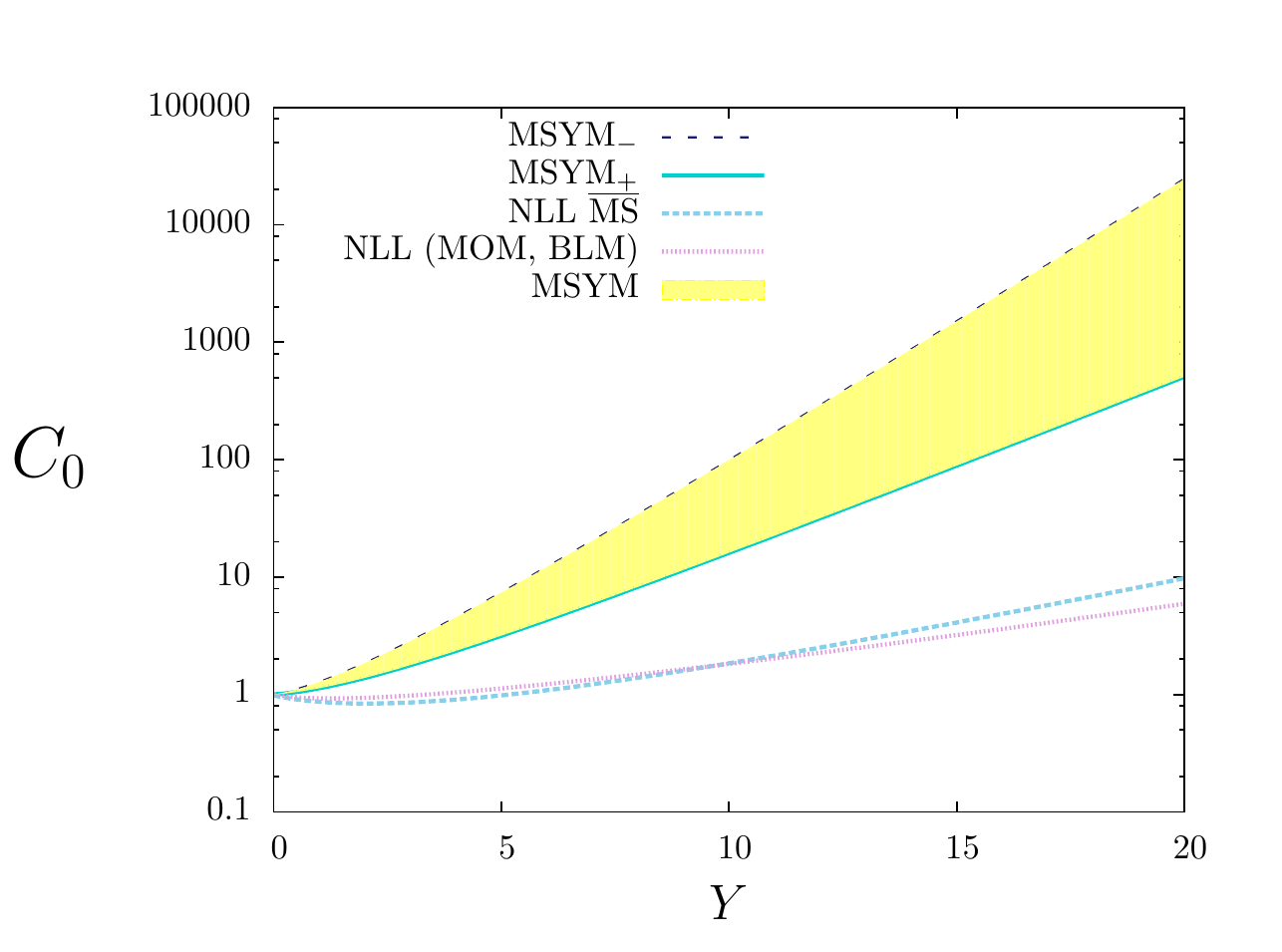}
\caption{Growth with dijet rapidity separation of the cross section in MSYM and QCD for different 
renormalization schemes.}
\label{u}
\end{figure}

As eq.~\eqref{total} shows, the $n=0$ coefficient is directly related to the normalized cross section:
$\hat{\sigma}(Y)/\hat{\sigma}(0) = \mathcal{C}_0(Y) / \mathcal{C}_0(0)$. The rise of $C_0$ with the rapidity separation between the two tagged jets is shown in \textsc{Fig.}~\ref{u}. The first feature to be noticed is the faster growth of the MSYM cross section for the wide range of values in the 't Hooft coupling, 
$\lambda$, considered in the yellow band. This shows that the NLO real emission kernel in the MSYM 
theory dominates over the virtual contributions in a much stronger fashion than in the QCD case, for any 
renormalization scheme. This also indicates that the net effect of introducing the extra fields in the supersymmetric 
multiplet increases the minijet multiplicity in the final state. In future works it will be worth looking into these 
details of the final state using event generator Monte Carlo techniques~\cite{Chachamis:2011rw}. 

It is noteworthy that for small rapidity separations the QCD cross section in the $\overline{\rm MS}$-scheme is lower than in the MOM-BLM 
scheme, with the latter being closer to the MSYM result. This is to be expected from a renormalization set-up which resums the higher order dominant conformal contributions. However, at $Y \simeq 10$ 
a departure from this behaviour is observed, with the calculation in the $\overline{\rm MS}$-scheme being 
now closer to the MSYM one. This again hints at the instability of the $n=0$ conformal spin component, which is 
known to be very sensitive to collinear radiation, not fully included in the BFKL kernel. 
\begin{figure}[htbp]
\centering
\includegraphics[scale=0.8]{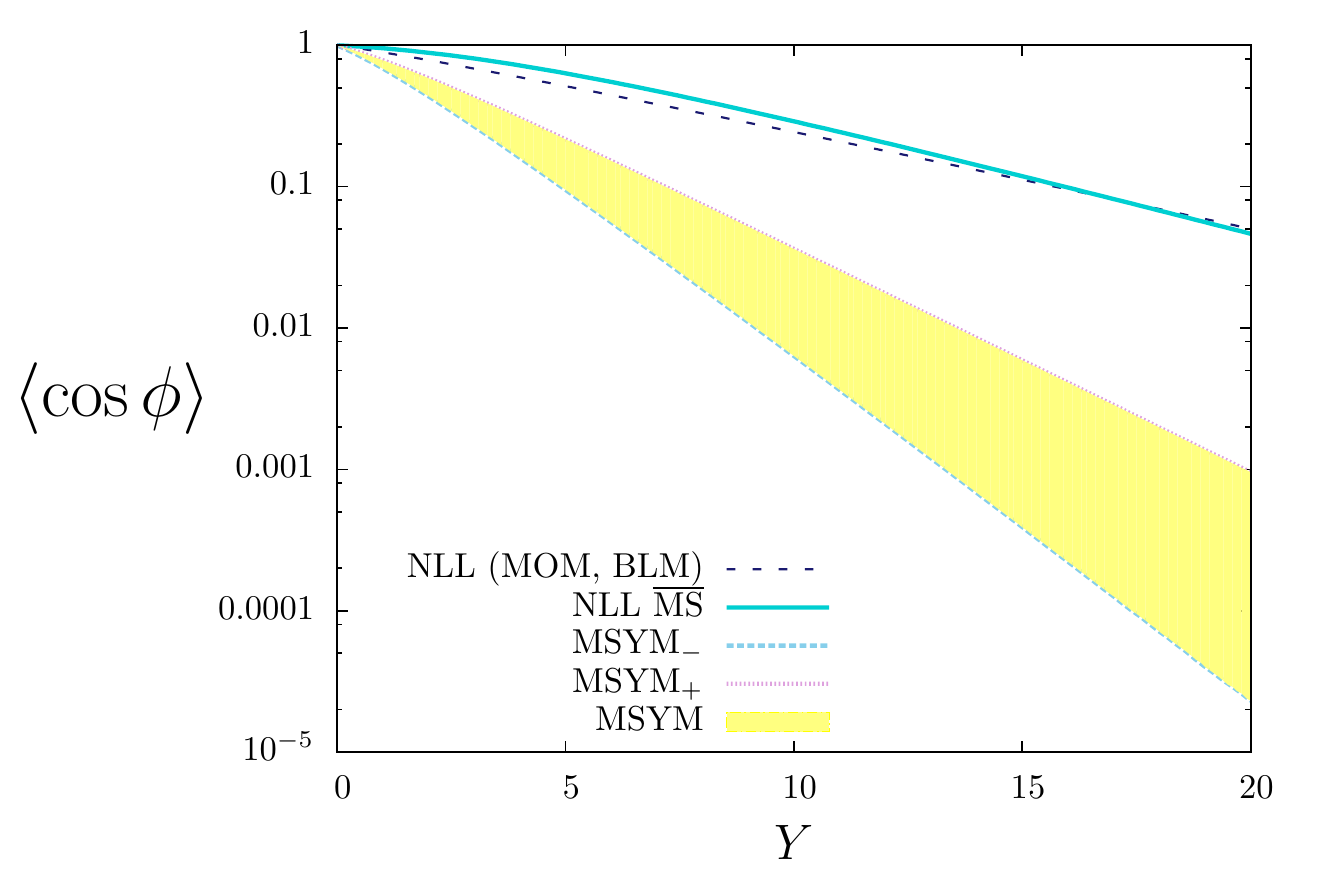}
\caption{Evolution of the average of  $\cos \phi $ with jet rapidity separation in MSYM and QCD for different renormalization schemes.}
\label{v}
\end{figure}
\begin{figure}[htbp]
\centering
\includegraphics[scale=0.8]{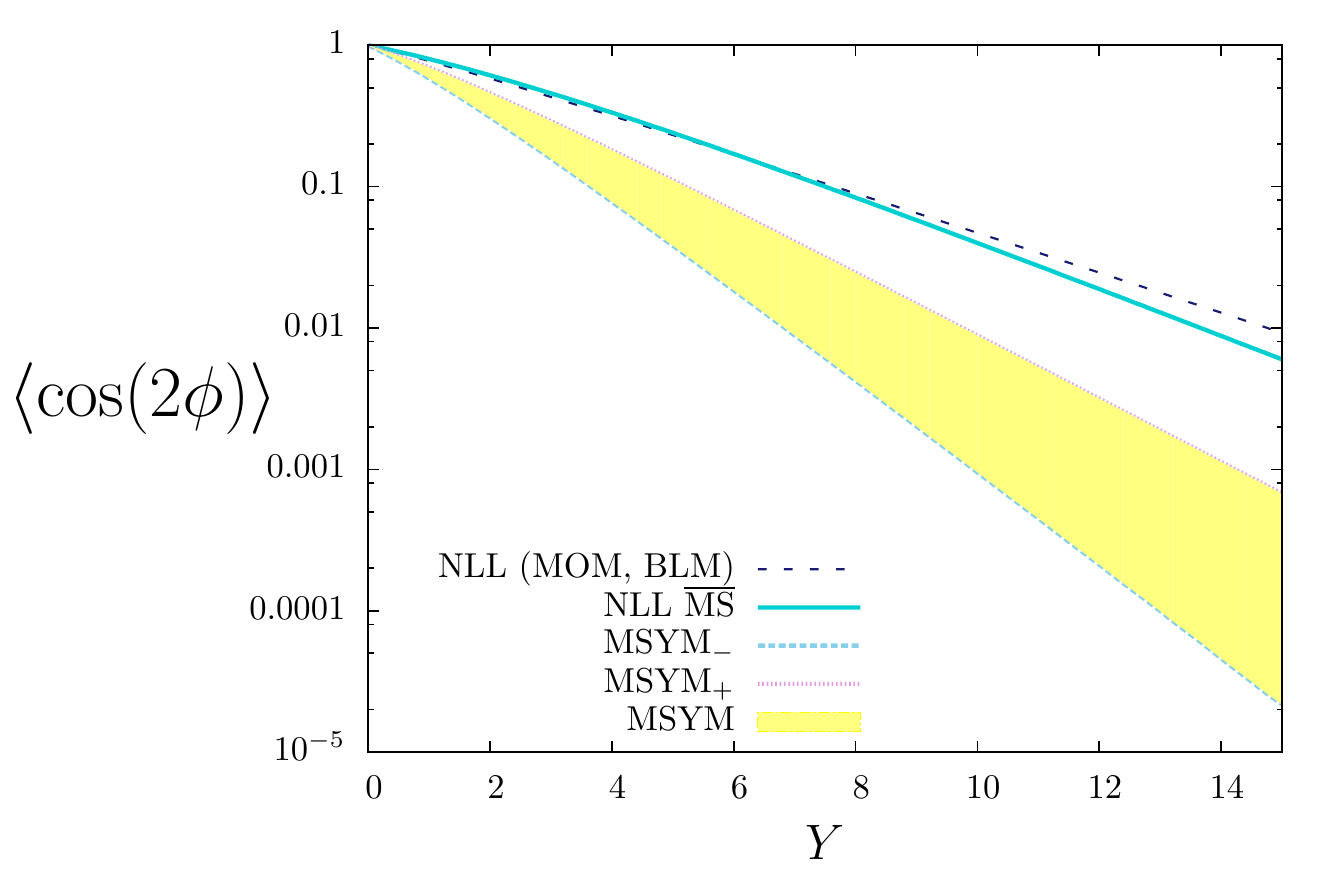}
\caption{Evolution of the average of  $\cos(2\phi)$ with jet rapidity separation in MSYM and QCD for different renormalization schemes.}
\label{w}
\end{figure}
It is then natural to 
predict a similar crossing of behaviour at some $Y$ for any quantity sensitive to the $n=0$ eigenvalue of the kernel. This is indeed what is found when the average of $\cos{\left( n \phi \right)}$ is calculated as 
in eq.~\eqref{cosmphi}. A couple of examples with a cross-over of lines are plotted in \textsc{Figs.} \ref{v} and \ref{w} for the $n=1,2$ cases. 
In these figures it is also interesting to note that the MSYM dijets are less correlated in azimuthal angle in the MSYM theory than in QCD, which corresponds to another manifestation of the higher multiplicity of parton radiation in the supersymmetric case. 
\begin{figure}[htbp]
\centering
\includegraphics[scale=0.8]{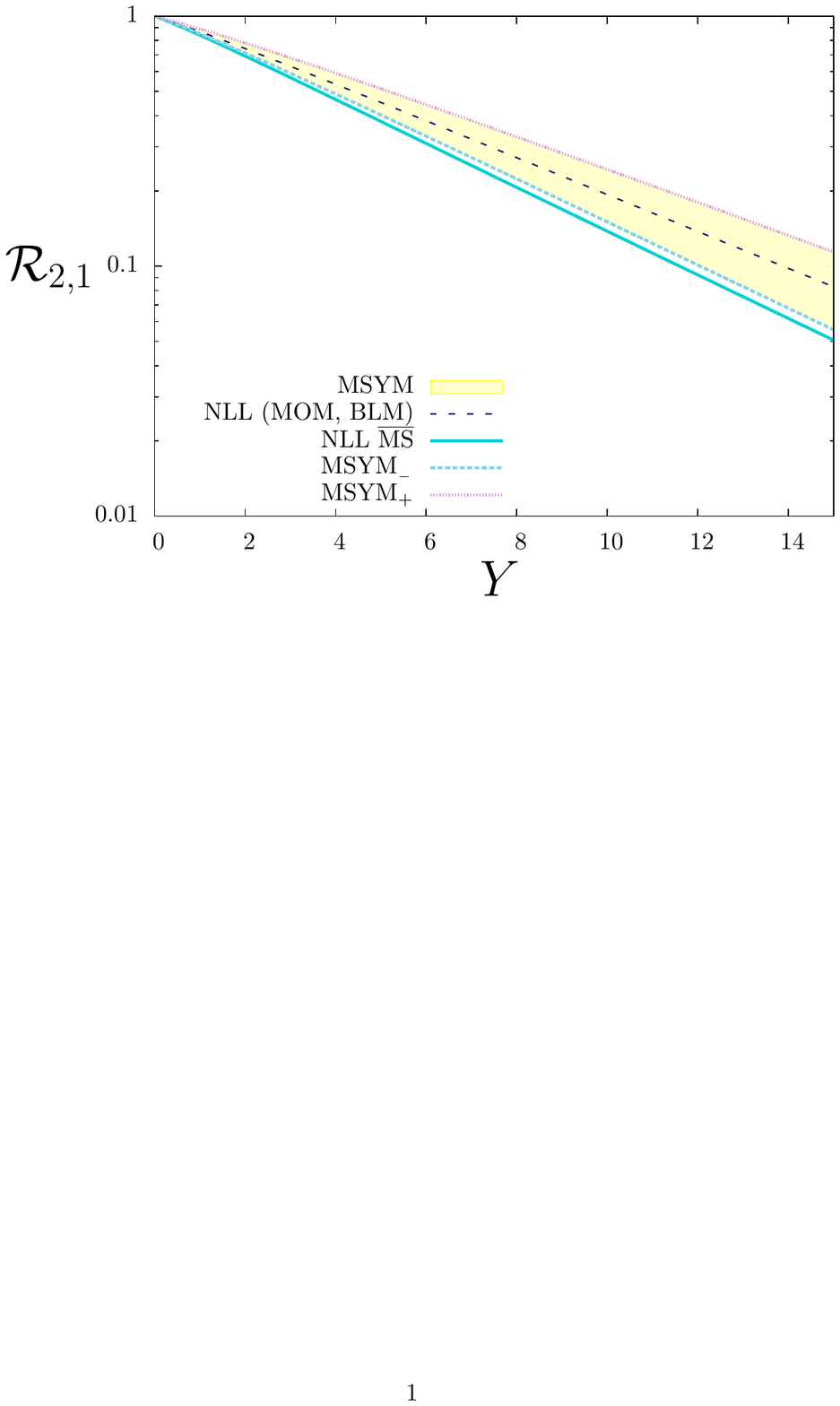}
\caption{Evolution of ${\cal R}_{2,1} = \frac{\langle\cos (2\phi)\rangle}{\langle \cos \phi\rangle}$ with jet rapidity separation in MSYM and QCD for different renormalization schemes.}
\label{x}
\end{figure}
\begin{figure}[htbp]
\centering
\includegraphics[scale=0.8]{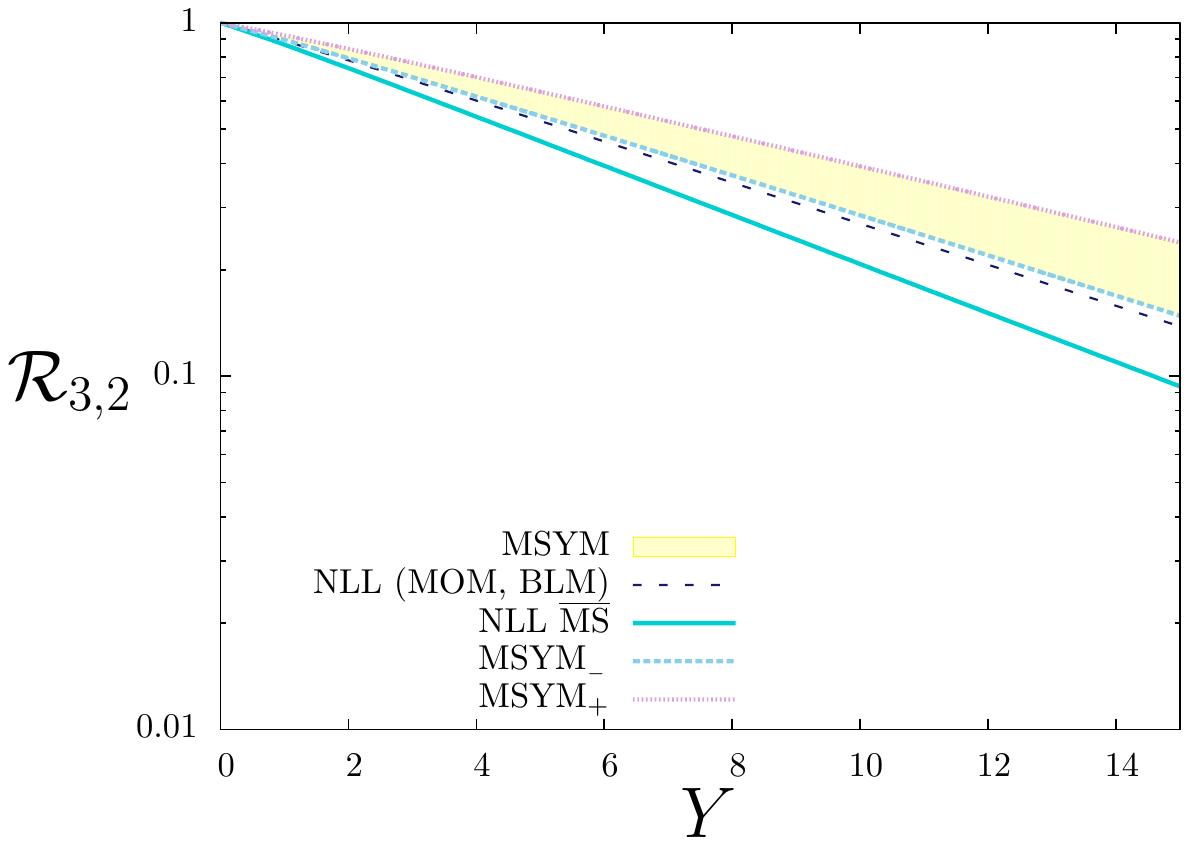}
\caption{Evolution of ${\cal R}_{3,2} = \frac{\langle\cos (3\phi)\rangle}{\langle \cos (2\phi)\rangle}$ with jet rapidity separation in MSYM and QCD for different renormalization schemes.}
\label{z}
\end{figure}

The main conclusion of this analysis is that to define observables only sensitive 
to conformal dynamics it is needed to remove the $n=0$ contribution. One way of doing this is to 
use the ratios of azimuthal angle averages ${\cal R}_{m,n} (Y)$ in eq.~\eqref{ratios}.  
Two of these ratios (${\cal R}_{2,1}$, ${\cal R}_{3,2}$) are calculated 
in \textsc{Figs.} \ref{x} and \ref{z}. It is important to note that the MSYM ratios are very close to those 
calculated in QCD, indicating that these observables capture the bulk of the conformal dynamics 
in QCD. Moreover, the renormalization scheme which gives ratios closest to those in MSYM is 
the MOM-BLM scheme, independently of the separation in rapidity between the two tagged jets. Having 
removed the $n=0$ dependence, the cross-over of lines does not take place anymore.

\section{Conclusions}\label{5}

In this work dijet production has been investigated for configurations where the two tagged jets are largely 
separated in rapidity. This has been done both in QCD and MSYM, with particular emphasis in investigating 
how different renormalization procedures in QCD best reproduce the conformal dynamics of MSYM. With this 
purpose, ratios of azimuthal angle decorrelations, which have an excellent perturbative convergence, have been shown to capture the bulk of the conformal contributions. For these ratios the results calculated in QCD with the MOM-BLM scheme are very similar to 
those obtained in MSYM. It has also been shown that the two tagged jets are less correlated in azimuthal 
angle in MSYM than in QCD, indicating that in the supersymmetric case there is a larger multiplicity in the 
final state. For future work it will be interesting to investigate other multijet configurations and different 
observables at a more exclusive level.\\\\
{\bf \large Acknowledgements}\\\\
M. A. and G. C. would like to thank the Department of Theoretical Physics at the Aut{\'o}noma University of Madrid and the ``Instituto de F{\'\i}sica Te{\' o}rica UAM / CSIC" for their hospitality. A. S. V. would like to thank the organizers of the KITP programs ``The Harmony of Scattering Amplitudes" and ``The First Year of the 
LHC" where the latest stages of this work have been completed. This research has been partially supported by the European Comission under contract LHCPhenoNet (PITN-GA-2010-264564) and the Comunidad de Madrid through Proyecto HEPHACOS ESP-1473.

\end{document}